\documentclass[aps,pre,preprint,showpacs]{revtex4-1}
\usepackage[usenames]{color}
\usepackage{amssymb,amsfonts,amsmath}

\def\mean#1{\langle #1 \rangle}

\begin{document}
\title{Decomposing Entropy Productions by Double Control Parameters}
\author{Jang-il \surname{Sohn}}
\email{physicon@korea.ac.kr}
\affiliation{Department of Physics, Korea University, Seoul 136-713, Korea}

\date{\today}

\begin{abstract}
In the present work, we study the entropy productions in a system controlled by double control parameters.
By introducing a thermal fluctuation part, we solve the problem that the second law of the thermodynamics seems to be violated by the thermal fluctuation near equilibrium in the microscopic levels.
Then we define the negative and the compensating entropy productions in the macroscopic levels.
\end{abstract}


\maketitle

\section{Introduction}

As very well known, the second law seems to be violated in the microscopic levels.
To solve the problems, various versions of the fluctuation theorems \cite{PhysRevLett.78.2690,PhysRevLett.89.050601,PhysRevLett.86.3463,oono1998steady,PhysRevE.60.R5017,speck2005integral,chernyak2006path,PhysRevLett.104.090601,seifert2012stochastic} have been studied ever since the first version of which was introduced by Evans, Cohen, and Morriss in 1993 \cite{PhysRevLett.71.2401}, and they have been verified through experiments (reviewed in \cite{seifert2012stochastic}).
Among them, the detailed fluctuation theorem is given by 
\begin{equation}
\label{detailed_FT}
\frac{P(+\Delta S)}{P(- \Delta S)} = e^{\Delta S},
\end{equation}
where $P(\pm \Delta S)$ is the probability of the entropy increasing ($+\Delta S$) or decreasing ($-\Delta S$) events \cite{PhysRevLett.71.2401,crooks1998nonequilibrium,PhysRevE.60.2721,PhysRevLett.104.090601,seifert2012stochastic}.
The relation explains the entropy productions in the nonequilibrium processes very well.
However, the relation can not explain the violation of the second law of the thermodynamics near equilibrium.
For an example, let us imagine a system fluctuating by the thermal fluctuation near equilibrium.
In the thermodynamic limits, $\Delta S \rightarrow 0$, the relation (\ref{detailed_FT}) does not matter.
However, in the microscopic levels, (\ref{detailed_FT}) could be violated because of the thermal fluctuation, ${P(+\delta S)} = {P(- \delta S)}$, where $\delta S$ is entropy change due to thermal fluctuation.
We think that the relation (\ref{detailed_FT}) is right in itself, but something has been missed.
As seen in ref. \cite{PhysRevLett.104.090601}, $\Delta S$ in (\ref{detailed_FT}) can be one of $\Delta S_{tot}$, $\Delta S_{na}$ and $\Delta S_{a}$.
Thus, the first purpose of this work is to explain the thermal fluctuation in $\Delta S_{tot/na/a}$ near equilibrium.
We will show that the entropy productions due to the thermal fluctuation are canceled out in $\Delta S_{na}$.

The second purpose of this work is to find the negative and the compensating entropy productions, which are associated with the so-called Schr\"odinger's paradox.
In the book {\it What is Life?} \cite{schrodinger1992life}, Schr\"odinger said that life feeds on negative entropy to avoid decay to equilibrium (or death).
His negative entropy seems to be violating the second law of thermodynamics in the macroscopic levels, which is the ``Shr\"odinger's Paradox''.
According to Prigogine, the negative entropy production is compensated by the positive entropy production due to the heat flow in reservoir \cite{prigogine1961introduction,prigogine1989entropy}.
Though his explanation is widely accepted, the negative and the compensating entropy productions were not discussed enough, so far.
In this work, we define the negative and the compensating entropy productions, and show that they are mutually compensating in nonequilibrium process.

In this work, we discuss the problems in a system controlled by the double control parameters, $\Lambda = (\lambda,\rho)$.
Since energy or Hamiltonian is generally not defined well in nonequilibrium system, we begin at the Boltzmann distribution (or the canonical ensemble), and expand it to the nonequilibrium processes introducing entropic distances $\phi$ and $\theta$, instead of directly approaching the nonequilibrium systems.

\section{Entropy Productions}
\label{sec:entropy_productons}

\subsection{Definitions and Assumptions}
\label{definitions}

We study the entropy productions in Markov processes such that
\begin{equation}
\begin{split}
p_{i}(t) &= \sum_{j}w_{ij}(\Lambda_{t}) p_{j}(t-1) 
\end{split}
\end{equation}
where $p_{i}(t)$ is the probability distribution of the state of the system $i$ at time $t$, $w_{ij}(\Lambda_{t})$ is the time evolution operator (or propagator) from a previous state $i$ to a next state $j$, where $\Lambda_{t} = (\lambda_{t},\rho_{t})$ is a set of control parameters at time $t$.
To separate the dynamics of each control parameter, we decompose $w_{ij}(\Lambda)$ logarithmically into two parts as follows
\begin{equation}
\label{eq:AS_operator}
\begin{split}
&w_{ij}(\Lambda) = \epsilon_{ij}(\lambda)  \nu_{ij}(\rho) \quad \textrm{or} \\
&\ln w_{ij}(\Lambda) = \ln \epsilon_{ij}(\lambda) + \ln \nu_{ij}(\rho)
\end{split}
\end{equation}
where $\epsilon_{ij}(\lambda)$ and $\nu_{ij}(\rho)$ are the non-driving and driving parts, respectively.

Between the control parameters, the former one $\lambda$ is associated with the {\it non-driving forces} which does not generate nonequilibrium flux in steady states.
The non-driving forces include all kinds of the conservative forces and the {\it non-conservative non-driving} forces ($e.g.$ the friction or the resistive forces).
So if the system was originally out of equilibrium while $\dot \lambda = 0$, the state of the system is attracted to the equilibrium state by the conservative forces, and damping by the non-conservative non-driving forces (the friction or the resistive forces), then eventually the system reaches the equilibrium state, provided that no the driving force is applied.
Thus, for instance, $\lambda$ can be the position of the piston in a gas system, the slowly varying electromagnetic fields, the displacement of optical tweezers,  $etc$.

The another control parameter, $\rho$, is associated with the {\it driving forces} generating the nonequilibrium flux, $e.g.$ the convection of fluid, the electric current in a circuit, the biochemical cycles in ecosystems, the chemical reaction cycles in organisms, the non-zero flux of particles due to Maxwell's Demon, $etc$.
In the absence of the driving forces, we can define $\rho^{eq}$ such that 
\begin{equation}
  \label{eq:detailed_balance_eq}
  \frac{w_{ij}(\lambda,\rho^{eq})p^{st}_{j}(\lambda,\rho^{eq})}{w_{ji}(\lambda,\rho^{eq})p^{st}_{i}(\lambda,\rho^{eq})} = 1,
\end{equation}
where $p^{st}_{i}(\lambda,\rho)$ is a steady state distribution for fixed $\Lambda=(\lambda,\rho)$.
Thus, we define the symmetric operator as
\begin{equation}
\epsilon_{ij}(\lambda) \equiv w_{ij}(\lambda,\rho^{eq}),
\end{equation}
and define the equilibrium distribution by Boltzmann distribution,
\begin{equation}
\label{Boltzmann_distribution}
p^{eq}(\lambda) \equiv p^{st}(\lambda,\rho^{eq}) = e^{\beta(F-E_i)},
\end{equation}
where $\beta= 1/k_B\mathcal{T}$ is the reciprocal temperature of the system, $E_i$ is the energy level of $i$'th state, and $F = -k_B \mathcal{T} \ln \sum_{i} e^{-\beta E_{i}}$ is the Helmholtz free energy.

Then, we define the driving factor as
\begin{equation}
\label{def:nu}
\nu_{ij}(\rho) \equiv \frac{w_{ij}(\lambda,\rho)}{\epsilon_{ij}(\lambda)},
\end{equation}
and assume that
\begin{equation}
\label{steady_distribution}
p^{st}_{i}(\lambda,\rho) = p^{eq}_{i}(\lambda) e^{-\theta_i},
\end{equation}
where $\theta_i$ is an {\it entropic distance} from $p^{eq}_{i}(\lambda)$ to $p^{st}_{i}(\Lambda)$ for a state $i$.
If $\rho=\rho^{eq}$, then
\begin{equation}
\label{same_with_equilibrium}
\begin{split}
&p^{st}_{i}(\lambda,\rho^{eq}) = p^{eq}_{i}(\lambda)
\quad(\textrm{or }\theta_{i}(\rho^{eq}) = 0)\\
&\textrm{and }
\nu_{ij}(\rho^{eq}) = 1
\end{split}
\end{equation}
by definition.
If an external force begins to drive a system, which was originally in an equilibrium state, to a nonequilibrium steady state by $\rho \ne \rho_{eq}$, the detailed balance condition is not satisfied.

Additionally, we introduce $\phi_i$ which indicates the {\it entropic distance} from $p^{st}_i(\lambda_t,\rho_t)$ to $p_{i}(t)$ at time $t$, $i.e.$ 
\begin{equation}
\label{probability_distribution}
p_{i}(t) = p^{st}_{i}(\lambda_t, \rho_t) e^{-\phi_i}.
\end{equation}
Here, non-zero $\phi_i$ is generated by thermal fluctuations or perturbations.
Hence, from (\ref{Boltzmann_distribution}) and (\ref{steady_distribution}), we can write the probability distribution as follows
\begin{equation}
\label{assumed_distribution}
p_{i}(t) = p^{st}_{i}(\Lambda_t) e^{-\phi_i} = p^{eq}_{i}(\lambda_t) e^{-\theta_i-\phi_i},
\end{equation}
where $\Lambda_t = (\lambda_t,\rho_t)$.

\subsection{Total Path Entropy Production}


When a system evolves along a path of the states $ [i] = \{i_0,\cdots,i_t,\cdots,i_T\}$ by a schedule of the control parameter 
$[\Lambda] = \{(\lambda_0,\rho_0), \cdots, (\lambda_t,\rho_t), \cdots, (\lambda_T,\rho_T) \},$
we write the path as a set of them as follows
$$[i;\Lambda] \dot= \left[ i_0 \xrightarrow {(\lambda_{1},\rho_{1})} i_{1} \xrightarrow {(\lambda_{2},\rho_{2})} \cdots \xrightarrow {(\lambda_{T},\rho_{T})} i_{T} \right],$$
where $\Lambda_0 = (\lambda_{0},\rho_{0})$ is the initial control parameter before the beginning of the schedule.
The probability to evolve along the path in the forward direction (path probability) is given by
\begin{equation}
\label{path_probability}
\mathcal{P}[i;\Lambda] = \prod_{t=1}^{T}w_{i_{t}i_{t-1}}(\Lambda_{t}) p_{i_0}(0).
\end{equation}
Along the reversed direction, $i_0 \xleftarrow {\Lambda_{1}} i_{1} \xleftarrow {\Lambda_{2}} \cdots \xleftarrow {\Lambda_{T}} i_{T}$, the reversed path probability is given by
\begin{equation}
\label{rpath_probability}
\bar{\mathcal{P}}[i;\Lambda] = \prod_{t=1}^{T} w_{i_{t-1}i_{t}}(\Lambda_{t}) p_{i_T}(T),
\end{equation}
where the over-bar indicates the reversed path.
If $\Delta S$ has been generated when the system evolves along the forward path, $-\Delta S$ will be generated in the reversed path.
Therefore, the detailed fluctuation theorem (\ref{detailed_FT}) can be directly applied as follows
\begin{equation}
\label{total_path_entropy}
\begin{split}
\Delta S_{tot} [i;\Lambda]
&\equiv \ln \frac{\mathcal{P}[i;\Lambda]}{\bar{\mathcal{P}}[i;\Lambda]} \\
&= \ln \prod_{t=1}^{T} \frac{w_{i_{t}i_{t-1}}(\Lambda_t)p_{i_0}(0)}{w_{i_{t-1}i_{t}}(\Lambda_t)p_{i_T}(T)}
\end{split}
\end{equation}
as shown by Crooks in ref. \cite{PhysRevE.60.2721}.

\subsection{Decomposition of Path Entropy Production}
\label{sec:decompositions}

The path probability along the forward path (\ref{path_probability}) can be decomposed
as follows
\begin{equation}
\label{eq:AS_path}
\ln \mathcal{P}[i;\Lambda] = \ln \mathcal{P}_{\epsilon}[i;\Lambda] + \ln \mathcal{P}_{\nu}[i;\Lambda],
\end{equation}
\begin{align*}
\textrm{where}\quad\mathcal{P}_{\epsilon}[i;\Lambda] \equiv \prod_{t=1}^{T} \epsilon_{i_{t}i_{t-1}}(\lambda_{t}) p_{i_0}(0) \\
\textrm{and}\quad\mathcal{P}_{\nu}[i;\Lambda] \equiv \prod_{t=1}^{T} \nu_{i_{t}i_{t-1}}(\rho_{t})
\end{align*}
since (\ref{eq:AS_operator}).
Note that $\mathcal{P}_{\nu}[i;\Lambda]$ is not a path probability but a {\it path-probability-like} quantity.
In the same manner, the reversed path probability (\ref{rpath_probability}) can be decomposed as follows
\begin{equation}
\label{eq:AS_path_r}
\ln \bar{\mathcal{P}}[i;\Lambda] = \ln \bar{\mathcal{P}}_{\epsilon}[i;\Lambda] + \ln \bar{\mathcal{P}}_{\nu}[i;\Lambda]
\end{equation}
\begin{align*}
\textrm{where}\quad\bar{\mathcal{P}}_{\epsilon}[i;\Lambda]= \prod_{t=1}^{T} \epsilon_{i_{t-1}i_{t}}(\lambda_{t}) p_{i_T}(T) \\
\textrm{and}\quad\bar{\mathcal{P}}_{\nu}[i;\Lambda] =\prod_{t=1}^{T} \nu_{i_{t-1}i_{t}}(\rho_{t}). 
\end{align*}
Thus, the total entropy production (\ref{total_path_entropy}) can be rewritten as
\begin{equation}
\label{eq:AS_entropy}
\Delta S_{tot}[i;\Lambda] = \Delta S_{\epsilon}[i;\Lambda] + \Delta S_{\nu}[i;\Lambda]
\end{equation}
where
\begin{align}
\label{symmetric_entropy}
&\Delta S_{\epsilon}[i;\Lambda] \equiv \ln \frac{\mathcal{P}_{\epsilon}[i;\Lambda]}{\bar{\mathcal{P}}_{\epsilon}[i;\Lambda]} =\ln \prod_{t=1}^{T} \frac{\epsilon_{i_{t}i_{t-1}}(\lambda_{t}) p_{i_0}(0)}{\epsilon_{i_{t-1}i_{t}}(\lambda_{t}) p_{i_T}(T)} \\
\label{asymmetric_entropy}
&\textrm{and}\quad
\Delta S_{\nu}[i;\Lambda] \equiv \ln \frac{\mathcal{P}_{\nu}[i;\Lambda]}{\bar{\mathcal{P}}_{\nu}[i;\Lambda]} = \ln \prod_{t=1}^{T} \frac{\nu_{i_{t}i_{t-1}}(\rho_{t})}{\nu_{i_{t-1}i_{t}}(\rho_{t})}.
\end{align}
Here, $\Delta S_{\epsilon}$ and $\Delta S_{\nu}$ can be decomposed again as follows
\begin{align}
\label{e_entropy}
\Delta S_{\epsilon}[i;\Lambda] = \Delta S_{sys}[i;\Lambda] + \Delta S_{h}[i;\Lambda] \\
\label{n_entropy}
\Delta S_{\nu}[i;\Lambda] = \Delta S_{res}[i;\Lambda] - \Delta S_{h}[i;\Lambda]
\end{align}
where
\begin{align}
\label{system_entropy}
&\Delta S_{sys}[i;\Lambda] \equiv \ln \frac{p_{i_0}(0)}{p_{i_{T}}(T)},\\
\label{reservoir_entropy}
&\Delta S_{res}[i;\Lambda] \equiv \ln \prod_{t=1}^{T} \frac{w_{i_{t}i_{t-1}}(\Lambda_t)}{w_{i_{t-1}i_{t}}(\Lambda_t)},  \quad \textrm{and}\\
\label{Delta_S_epsilon-ex}
&\Delta S_{h}[i;\Lambda]
\equiv \ln \prod_{t=1}^{T} \frac{\epsilon_{i_{t}i_{t-1}}(\lambda_{t})}{\epsilon_{i_{t-1}i_{t}}(\lambda_{t})}
=\ln \prod_{t=1}^{T} \frac{p^{eq}_{i_{t}}(\lambda_{t})}{p^{eq}_{i_{t-1}}(\lambda_{t})}.
\end{align}
Here, $\Delta S_{h}[i;\Lambda]$ is the entropy production due to heat as seen in (\ref{eq_excess_entropy}) (the proof is written in Appendix \ref{Equilibrium_Excess_Entropy_Production}).



\subsection{Entropy Production Rates}
\label{sec:entropy_production_rates}

In the continuous time limits, (\ref{e_entropy}) and (\ref{n_entropy}) become 
\begin{align}
\label{eq:epsilon_derivative}
\frac{d {S_{\epsilon}}}{dt}
&=\frac{d {S_{sys}} }{ d t } + \frac{d S_{h}}{d t} \\
\label{eq:nu_derivative}
\frac{d {S_{\nu}}}{dt} 
&= \frac{d {S_{res}}}{dt} - \frac{d S_{h}}{d t}
\end{align}
where 
\begin{align}
\label{system_entropy_production_rate}
&\frac{d {S_{sys}} }{dt}
= \frac{d S}{dt} + \frac{d \phi}{ dt } + \frac{d \theta}{ dt} \\
\label{eq_excess_entropy}
&\frac{d { S_{h}}}{dt}  = -\beta \frac{d Q }{d t}.
\end{align}
Here, $\phi \equiv \sum_{i}p_{i}(t) \phi_i$ and $\theta \equiv \sum_{i}p_{i}(t) \theta_i$ are the average entropic distances of the system, $Q = E - W $ is the heat into the system, and 
\begin{equation}
\label{def:S_eq}
S \equiv -\beta F + \beta {E}
\end{equation}
is the entropy relevant to the Helmholtz free energy $F$ and the internal energy $E \equiv \sum_{i}p_{i}(t){E_{i}}$ in the equilibrium physics.

Additionally, from the three detailed FTs \cite{PhysRevLett.104.090601}, we can see that
\begin{align}
\label{ineq:na_derivative}
\frac{d {S_{na}}}{dt} 
&= \frac{d {S_{\epsilon}}}{dt} - \frac{\partial {\theta}}{\partial t} \ge 0 \\
\label{ineq:a_derivative}
\frac{d {S_{a}}}{dt}
&= \frac{d {S_{\nu}}}{dt} + \frac{\partial {\theta}}{\partial t} \ge 0.
\end{align}
because
\begin{align}
\label{excess_relation}
\frac{d S_{ex}}{d t} &= \frac{d{S_{h}}}{dt} - \frac{\partial {\theta}}{\partial t}, \\
\Delta S_{na} &= \Delta S_{sys} + \Delta S_{ex}, \\
\textrm{and}\quad\Delta S_{a} &= \Delta S_{res} - \Delta S_{ex}.
\end{align}

The three integral or detailed FTs are briefly reviewed in {\bf Appendix \ref{Brief_Review}}, and the proofs of (\ref{system_entropy_production_rate}$\sim$\ref{excess_relation}) are written in {\bf Appendix \ref{sec:derivative}}.

\section{The Second Law of Thermodynamics}
\label{sec:second_law_fluctuation}

\subsection{Thermal Fluctuation Near Equilibrium}
\label{sec:thermal_fluctuation}

In this section, we study the entropy productions in the absence of the driving forces, $\rho = \rho^{eq}$.
In the cases, 
the total entropy production is given by
\begin{align}
\label{system_entropy_equilibrium_cases_1}
&\frac{d {S_{tot}}}{dt}\bigg|_{\rho^{eq}} 
= \bigg[\frac{d {S_{\epsilon}}}{dt}
+ \frac{d{S_{\nu}}}{dt}\bigg]_{\rho^{eq}}
= \frac{d {S_{\epsilon}}}{dt}\bigg|_{\rho^{eq},\dot \beta = 0} \nonumber \\
&= \frac{d S}{dt} + \frac{d { \phi }}{dt} + \frac{d S_{h}}{dt} \nonumber \\
&= \frac{\partial S}{\partial \lambda}\dot \lambda + \frac{\partial S}{\partial \beta}\dot \beta + \frac{\partial S}{\partial t} + \frac{d { \phi }}{dt} - \beta \frac{d Q}{d t} 
\ge 0
\end{align}
since (\ref{eq:epsilon_derivative}) and (\ref{eq_excess_entropy}).
So if the system is fluctuating near equilibrium, then
\begin{equation}
\label{d_phi_d_Q_zero}
\frac{d { \phi }}{dt}\bigg|_{eq} - \beta \frac{d Q}{d t}\bigg|_{eq}= 0,
\end{equation}
because $\dot \lambda =0$, $\dot \beta= 0$, and $\frac{\partial S}{\partial t}\big|_{eq} = \beta \frac{\partial F}{\partial t}\big|_{eq}-\beta\frac{\partial E}{\partial t}\big|_{eq}=0$ (see {\bf Appendix \ref{F_E_eq}}) in equilibrium states.
Hence, we can define the entropy production due to the thermal fluctuation as
\begin{equation}
\label{def:thermal_fluctuation}
\delta S \equiv \frac{d \phi}{ d t}
\end{equation}
in the microscopic levels.
If $\delta S$ is given by an uncorrelated random variable,
$\overline{\delta  S} = 0$ and $\overline{ \delta S(t) \delta S(t')} = 2 \Phi \delta(t-t')$,
then the $rms$ of the thermal fluctuation $\delta Q \equiv - { \delta S}/{\beta} $ is
\begin{equation}
\label{fluctuation_heat_strength}
{\delta Q}_{rms} = \sqrt{2 \Phi} k_B \mathcal{T},
\end{equation}
which is consistent with the fluctuation-dissipation theorem, where the over-line refers to the time average, $\delta(t)$ is the delta function, and $\Phi$ is the strength of the fluctuation.
Therefore, replacing ${d\phi}/{dt}$ with $-\beta \delta Q$, the system entropy production rate (\ref{system_entropy_production_rate}) can be rewritten as follows
\begin{equation}
\label{system_entropy_production_rate_1}
\frac{d {S_{sys}} }{dt}
= \frac{d {S}}{d t} - \beta \delta Q + \frac{d \theta}{ dt}.
\end{equation}
Therefore, we claim that the violation of the second law in the microscopic levels can be fixed by considering $\delta Q$:
if a small system is in an equilibrium state, then the entropy production rate is given by identically zero,
\begin{equation}
\label{thermal_fluctuation_eq}
\begin{split}
\frac{d S_{tot}}{d t}\bigg|_{eq}
&= \left[ \underbrace{- \beta \delta Q}_{system} \right]_{eq}+ \left[\underbrace{-\beta \frac{d Q}{d t}}_{reservoir} \right]_{eq}
= 0,
\end{split}
\end{equation}
which states that the entropy production due to the thermal fluctuation in the system is equally compensated in the reservoir.
If $\delta Q$ is ignored, or if only system entropy is considered, the second law of thermodynamics is violated due to the thermal fluctuation in the microscopic levels.
But $\delta Q$ can be ignored in the thermodynamics limits.

\subsection{The Second Law in Isothermal Processes}
\label{subsec:isothermal}

In the thermodynamic limits, $\delta Q \simeq 0$, we can simply derive the ordinary (isothermal) second law of thermodynamics,
\begin{equation}
\label{isothermal_second_law}
\begin{split}
\frac{d S_{tot}}{d t}\bigg|_{\rho^{eq},\dot \beta = 0}^{\delta Q \simeq 0 }
&=\frac{d S_{\epsilon}}{d t}\bigg|_{\rho^{eq},\dot \beta = 0}^{\delta Q \simeq 0 } \\
&\simeq \frac{d S}{d t}\bigg|_{\dot \beta = 0} -\beta \frac{d Q}{dt} \\
&=-\beta \frac{d F}{dt} + \beta \frac{d W}{d t} \ge 0
\end{split}
\end{equation}
because $dE = dQ+dW=0$ in isothermal processes, where $W$ is work done {\it on system}.
Above inequality yields the well-known isothermal second law,
\begin{equation}
\label{isothermal_second_law_1}
W \ge \Delta F \quad\textrm{or} \quad W_{diss} \ge 0 ,
\end{equation}
where $W_{diss} = W - \Delta F$ is the dissipated work due to the non-conservative non-driving forces, $e.g.$ friction, resistive forces, $etc$.

\subsection{Negative Entropy Production}
\label{sec:negative}

In the {\it out-of-equilibrium processes}, the negative entropy can be produced in the system.
Assume that a system is driven out of equilibrium,
\begin{equation}
  \label{assume_out}
\begin{split}
p_{i}(0)=p^{st}_{i}(\lambda&,\rho^{eq}) = p^{eq}_{i}(\lambda) \\
&\Downarrow {[\Lambda_{out}]} \\
p_{i}(T)=p^{st}_{i}(\lambda&,\rho^{far}) \ne p^{eq}_i(\lambda),
\end{split}
\end{equation}
by a schedule
$$[\Lambda_{out}] = \{ (\lambda, \rho^{eq}), \cdots, (\lambda, \rho_t), \cdots, (\lambda, \rho^{far}) \},$$
where $\rho_0 = \rho^{eq}$, $\rho_{t\ne 0} \ne \rho^{eq}$ and $\rho_T=\rho^{far}$ for $t \in \{0,\cdots,T\}$, and fixed $\lambda$.
The super-script ``${far}$'' refers to ``{\it far from equilibrium}".
In the processes, we assume that $p_i(t) = p_i^{st}(\lambda, \rho_{t})$ ignoring thermal fluctuation, $\delta Q=0$.
After some algebra (written in {\bf Appendix \ref{derive:negative_entropy}}),  (\ref{symmetric_entropy}) yields an inequality,
\begin{align}
  \label{negative_epsilon}
\Delta S_{\epsilon} [\Lambda_{out}]
&= - D \Big( p^{st}_{i}(\Lambda_T) \Big| p^{eq}_{i}(\lambda) \Big) \le 0.
\end{align}
The above inequality is not weird, because negative entropy is generated in the system, whenever a system is driven out of equilibrium.

The relation $p^{st}_{i}(\Lambda) = p^{eq}_{i}(\lambda) e^{-\theta_i}$ yields an equality,
\begin{align}
\label{entropic_distance}
\sum_{i} p^{eq}_{i}(\lambda) = \sum_{i} e^{\theta_i}p^{st}_{i}(\Lambda) = \sum_{i} e^{\theta_i}p_{i}(t) \nonumber \\
\quad\rightarrow \quad  \mean{e^{\theta}}  = 1
\end{align}
since $p_{i}(t)=p^{st}_{i}(\Lambda)$ in this case.
By applying the Jensen's inequality \cite{ross1996stochastic} to (\ref{entropic_distance}), it is simply shown that the average entropic distance from the original equilibrium state is non-positive, 
\begin{equation}
\label{theta_inequality}
 \theta  \le  0.
\end{equation}
If $\rho = \rho^{eq}$, the entropic distance is maximized, $\theta = 0$, by definition.
Therefore, we can see that the entropic distance between the (driven) nonequilibrium steady state and the equilibrium state, $\theta_{i} = \ln \frac{p^{eq}_i (\lambda) }{ p^{st}_{i} (\Lambda) }$, is the negative entropy.
For the reasons, we define the negative entropy production along a path as follows
\begin{equation}
\begin{split}
\label{def:negative_entropy}
\Delta S_{neg}[i;\Lambda] &\equiv
\ln \frac{p^{eq}_{i_T}(\lambda_T)}{p^{st}_{i_T}(\Lambda_T)} - \ln\frac{p^{eq}_{i_0}(\lambda_0)}{p^{st}_{i_0}(\Lambda_0)} 
=\theta_{i_T} - \theta_{i_0}.
\end{split}
\end{equation}
Note that $\Delta S_{neg}[i;\Lambda]$ is path-independent.
Therefore, 
\begin{equation}
\label{def:negative_entropy_1}
  \frac{d S_{neg}}{dt} = \frac{d \theta}{d t}
  \quad \textrm{or} \quad
  \Delta S_{neg}[\Lambda] = \Delta \theta.
\end{equation}

Though we named it the {\it negative entropy production}, $\Delta S_{neg}$ is can be positive depending on the schedule of the control parameters.
For instance, in the into-equilibrium processes,
\begin{align}
p_{i_0}(0) = p^{st}_{i_0}&(\Lambda_0) \ne p^{eq}_{i_0}(\lambda_0) \nonumber \\
&\Downarrow {[\Lambda_{in}]} \\
p_{i_T}(T) = p^{st}_{i_T}&(\Lambda_T) = p^{eq}_{i_T}(\lambda_T), \nonumber
\end{align}
it has positive value,
\begin{equation}
\label{into_equilibrium}
\begin{split}
\Delta S_{neg}[\Lambda_{in}] 
&= - \sum_{i_T} p^{st}_{i_0}(\Lambda_0) \ln \frac{p^{eq}_{i_0}(\lambda_0)}{p^{st}_{i_T}(\Lambda_T)} \\
&= D\left( p^{st}_{i_T}(\Lambda_T) | p^{eq}_{i_T}(\lambda_T) \right) \ge 0,
\end{split}
\end{equation}
where $[\Lambda_{in}] = \{ (\lambda, \rho^{far}), \cdots, (\lambda, \rho_t), \cdots, (\lambda, \rho^{eq}) \}$.
We named $\Delta S_{neg}$ the negative entropy production in the sense that it is negative when the system is driven out of equilibrium by the driving forces.

\subsection{Compensating Entropy Production}
\label{subsec:compensating}

According to Prigogine, the negative entropy is compensated by the entropy productions in reservoir \cite{prigogine1961introduction}.
The explanation is widely accepted.
We find the compensating entropy in this subsection.

In the out-of-equilibrium processes, from (\ref{ineq:na_derivative}) and (\ref{negative_epsilon}), we can see that $-\int_{out} \frac{\partial {\theta}}{\partial t}dt \ge 0$.
On the other hand, from (\ref{ineq:na_derivative}) and (\ref{into_equilibrium}), we can see that $- \int_{in} \frac{\partial {\theta}}{\partial t}dt \le 0$ in the into-equilibrium processes.
Thus, we define compensating entropy production as
\begin{equation}
\label{compensating_theta}
\begin{split}
\frac{d {S_{c}}}{dt} \equiv - \frac{\partial {\theta}}{\partial t}
= \frac{d {S_{ex}}}{dt} - \frac{d {S_{h}}}{dt}
\end{split}
\end{equation}
in the continuous time limit.
Since $\frac{d S_{ex}}{dt} = -\beta \frac{d Q_{ex}}{dt}$ \cite{oono1998steady,PhysRevLett.86.3463,PhysRevLett.104.090601} and $\frac{d S_{h}}{dt} = -\beta \frac{dQ}{dt}$ in (\ref{eq_excess_entropy}), we can express the  compensating entropy production as
\begin{equation}
\label{compensating_entropy}
\frac{d {S_{c}}}{dt}
= - \beta \frac{d Q_{c}}{dt},
\end{equation}
where $Q_{c} \equiv Q_{ex} - Q$ is the Prigogine's compensating heat flowing out of the system in out-of-equilibrium processes.
From (\ref{Delta_S_epsilon-ex}), (\ref{ex_entropy}) and (\ref{compensating_theta}), the compensating entropy production along a path can be defined by
\begin{equation}
\label{identity_of_compensating_entropy_def}
\Delta S_{c} [i;\Lambda] \equiv \ln \prod_{t=1}^{T} \frac{p^{st}_{i_{t}}(\Lambda_t)p^{eq}_{i_{t-1}}(\lambda_{t})}{p^{st}_{i_{t-1}}(\Lambda_t)p^{eq}_{i_{t}}(\lambda_{t})}.
\end{equation}

\section{Conclusions and Discussions}

\subsection{Excess and House-Keeping Entropy Productions}

From (\ref{ineq:na_derivative}), (\ref{ineq:a_derivative}) and (\ref{compensating_entropy}), we can derive the excess part ($\Delta S_{na}$) and the house-keeping part ($\Delta S_{a}$) of the entropy production rates,
\begin{align}
\label{na_derivative_comp}
&\frac{d {S_{na}}}{dt} 
= \underbrace{\frac{d S_{sys}}{dt} + \frac{d {S_{h}}}{dt}}_{\frac{d {S_{\epsilon}}}{dt}}
+ \frac{d {S_{c}}}{dt} \ge 0\quad \textrm{and} \\
\label{a_derivative_comp}
&\frac{d {S_{a}}}{dt} = \underbrace{\frac{d {S_{res}}}{dt} - \frac{d {S_{h}}}{dt}}_{=\frac{d {S_{\nu}}}{dt}} - \frac{d {S_{c}}}{dt} \ge 0,
\end{align}
respectively, where
\begin{equation}
\label{general_theory:tot}
\frac{d {S_{tot}}}{dt} = \frac{d {S_{na}}}{dt} + \frac{d {S_{a}}}{dt} \ge 0.
\end{equation}
The inequality (\ref{na_derivative_comp}) states that $\Delta S_{c}[\Lambda]_{out}=-\beta Q_{c} \ge 0$ compensates $\Delta S_{neg}[\Lambda]_{out} \le 0$ in the system entropy $\Delta S_{sys} \simeq \Delta S + \Delta S_{neg}$ (ignoring $\delta Q$) in the out-of-equilibrium processes.
On the other hand, in the into-equilibrium processes, $\Delta S_{neg}[\Lambda]_{out} \ge 0$ compensates $\Delta S_{c}[\Lambda]_{out} \le 0$.
Therefore, $\Delta S_{neg}$ and $\Delta S_{c}$ are mutually compensating.

\subsection{Total Entropy Production}

From (\ref{nu_derivative}), (\ref{compensating_theta}), (\ref{a_derivative_comp}) and (\ref{compensating_entropy}), the reservoir entropy production rate is given by
\begin{equation}
\label{S_res_Q_tot}
\frac{d S_{res}}{d t} = - \beta \frac{d Q_{tot}}{dt} = - \beta \frac{d Q}{dt} - \beta \frac{d Q_{c}}{dt} - \beta \frac{d Q_{hk}}{dt}
\end{equation}
where the total heat flow is $dQ_{tot} = dQ + dQ_{c} + dQ_{hk}$, 
and $dQ_{ex} = dQ + dQ_{c}$.
Therefore, from the results so far,
the total entropy production can be expressed as follows, 
\begin{equation}
\label{total_result}
\begin{split}
&\frac{d S_{tot}}{dt} \\
&=\underbrace{\frac{d S}{d t} - \beta \delta Q - \frac{d S_{neg}}{dt}}_{\frac{d S_{sys}}{dt}} -\underbrace{ \beta \frac{d Q}{dt} - \beta \frac{d Q_{c}}{dt} - \beta \frac{d Q_{hk}}{dt}}_{\frac{d S_{res}}{d t}} \\
&=\underbrace{\frac{d S}{d t} - \beta \delta Q - \frac{d S_{neg}}{dt}- \beta \frac{d Q}{dt}}_{\frac{d S_{\epsilon}}{dt}} - \underbrace{ \beta \frac{d Q_{c}}{dt} - \beta \frac{d Q_{hk}}{dt}}_{\frac{d S_{\nu}}{d t}} \\
&=\underbrace{\frac{d S}{d t} - \beta \delta Q - \frac{d S_{neg}}{dt}- \beta \frac{d Q}{dt} - \beta \frac{d Q_{c}}{dt}}_{\frac{d S_{na}}{dt}} - \underbrace{ \beta \frac{d Q_{hk}}{dt}}_{\frac{d S_{a}}{d t}} \\
&\ge 0
\end{split}
\end{equation}
where $dS=-dF+dE$ is the entropy production in the thermodynamics, $\beta \delta Q = d \phi/dt$ is the entropy production due to the thermal fluctuation, $d S_{neg} = d \theta$ is the negative entropy production due to the driving forces, $d Q /d t = dE/dt -d W/dt$ is the heat into the system relevant to the first law of the thermodynamics, $dQ_{c} / d t = -\frac{\partial \theta}{\partial t}$ is the heat compensating the negative entropy production, and $dQ_{hk}/dt = dQ_{tot}/dt-dQ_{ex}/dt = dQ_{tot}/dt -dQ/dt - dQ_{c}/dt$ is the house-keeping heat due to the driving forces.

\subsection{Discussions}
\label{summary_discussions}


In the present work, we discussed about the violations of the second law of thermodynamics in both of the microscopic and the macroscopic levels.
We introduced the entropy production due to the thermal fluctuation, $\delta S= - \beta \delta Q$, in the microscopic levels, and defined the negative and the compensating entropy productions, $\Delta S_{neg}$ and $\Delta S_{c}$, in the macroscopic levels.

We studied the thermal fluctuation under the assumption that $\dot \lambda = 0$ and $\dot \beta = 0$ near equilibrium.
However, $\lambda$ and $\beta$ are actually fluctuating near equilibrium.
If the temperature of the system are extremely low, the fluctuation of temperature should be considered, $\dot \beta \rightarrow \delta \beta$.
Also, if the size of the system is very small, the fluctuation of work should be considered.
For instance, if a few particles of gas are enveloped by a chamber and a piston, the position of the piston can be fluctuating by the impacts of the particles, which can be called the work fluctuation, $\delta W = \left[\frac{\partial (-F + E)}{\partial \lambda } \right]_{eq} \delta \lambda$.
Thus, the fluctuating part can be given by
\begin{equation}
\delta S \equiv \frac{d \phi}{d t} = \frac{\delta W}{k_B \mathcal{T}} + \frac{ F - E }{k_B\mathcal{T}^2} \delta{\mathcal{T}} - \frac{\delta Q}{k_B \mathcal{T}}
\end{equation}
considering the fluctuation of $W$, $Q$ and $\mathcal{T}$.
Here, we can conjecture that why the second law of thermodynamics seems to be violated in the low-temperature systems \cite{PhysRevE.66.036102}: the entropy fluctuation is amplified, as temperature is decreasing, because $\delta S \sim \mathcal{T}^{-2}$.

About the thermal fluctuation, one may tackle the equality (\ref{d_phi_d_Q_zero}), and claim that it should be $\frac{d \phi}{dt}\big|_{eq} - \beta \frac{d Q}{dt}\big|_{eq} \ge 0$.
If the inequality form is right, the entropy is only growing even in the equilibrium states, then it should be $p_{i} = p^{eq}_{i}$ (or $\phi = 0$), and which implies no thermal fluctuation in equilibrium states.
However, the thermal fluctuations actually exist, indeed, unless in the ideal cases.
Therefore, the equality (\ref{d_phi_d_Q_zero}) is right.

After the discussions about the thermal fluctuation and the isothermal second law of the thermodynamics, we studied the negative and the compensating entropy productions in macroscopic levels.
We already know that if a system is driven out of an equilibrium state to a nonequilibrium steady state by the driving forces, the system entropy is reduced.
According to Prigogine, the negative entropy production (or reduced system entropy) is compensated by the positive entropy production by the heat flow in reservoir \cite{prigogine1961introduction}.
In the present work, we defined the negative and the compensating entropy productions, $\Delta S_{neg}$ and $\Delta S_{c}$, and found that they are mutually compensating.

Additionally, through (\ref{total_result}), we could understand that 1) how thermal fluctuation can be overcomed, and 2) why the three detailed (or integral) FTs \cite{PhysRevE.76.031132,PhysRevLett.104.090601}, $\Delta S_{tot/na/a} \ge 0$, are valid:
1) the thermal fluctuation $\delta Q$ is canceled out by $d Q/dt$ in steady states in $d S_{\epsilon} / d t$, so $\Delta S_{\epsilon/\nu}$ or $\Delta S_{na/a}$ are free from the thermal fluctuation;
2) the negative entropy production $d S_{neg}/dt$ is compensated by $\beta d Q_{c}/dt$ in $d S_{na}/ d t$, thereby the three integral or detailed fluctuation relations are satisfied in any situation.

If there is no driving force, the system goes to an equilibrium steady state, in which the detailed balance condition (\ref{eq:detailed_balance_eq}) is satisfied.
However, in the nonequilibrium steady states, the detailed balance condition is not satisfied.
Thus, from (\ref{eq:detailed_balance_eq}) and (\ref{probability_distribution}), we propose a generalized detailed balance condition as follows
\begin{equation}
\label{generalized_detailed_balance}
\frac{\epsilon_{ij}(\lambda) p_{j}(t)}{\epsilon_{ji}(\lambda) p_{i}(t)} = e^{(\phi_{j}-\phi_{i}) + (\theta_{j} - \theta_{i})},
\end{equation}
where $\phi$ and $\theta$ are the thermal fluctuation and the negative entropy, respectively, as seen in (\ref{def:thermal_fluctuation}) and (\ref{theta_inequality}).
The generalized detailed balance condition (\ref{generalized_detailed_balance}) implies that the system is constantly attracted to the equilibrium state by the conservative force to increase the system entropy (because $\epsilon_{ij}(\lambda)$ includes the conservative forces).
If the system is perturbed by the thermal fluctuation, the non-zero $\phi$ is generated, and which is dispersed by the friction or the resistive forces, and will be perturbed again, and so on.
As a result, the state of the system (or $\phi$) will be fluctuating near equilibrium (or zero).
If a constant driving force is acting on the system, $\nu_{ij}(\rho\ne\rho^{eq}) \ne 1$, the system is driven out of the original equilibrium steady state generating the negative entropy.
After the system have been reached the nonequilibrium steady state, it will be maintained by the house-keeping heat \cite{oono1998steady,PhysRevLett.86.3463,PhysRevLett.104.090601,seifert2012stochastic,PhysRevE.60.R5017} which is transformed to the reservoir entropy (or heat) by the friction or the resistive forces generally, or transformed to the other forms of energy, $e.g.$ electricity, photons, chemical energy, $etc$.

At last, to understand the {\it Schr\"odinger's negative entropy} completely, the relations between $\Delta S_{neg}$ and chemical potential $\mu$ should be studied further.
Considering the grand canonical ensemble, it can be conjectured that when a system is driven out of the original equilibrium state, $\Delta S_{neg}$ is generated, and which is transformed to the chemical energy through chemical reactions, $e.g.$ the photosynthesis in plants.
We expect that the transforming negative entropy is the Schr\"odinger's negative entropy feeding living things or organisms, and which can be expressed in the Gibbs free energy difference.

\section*{Acknowledgement}

We would like to appreciate Prof.s K.-I. Goh and I.-m. Kim in Korea University.
This works is supported by Basic Science Research Program through NRF grant 
funded by MEST (No. 2011-0014191).

\section*{Appendix}
\begin{appendix}

\section{. Brief Review of Three Integral or Detailed Fluctuation Theorems}
\label{Brief_Review}


Let us briefly review the so-called three integral or detailed fluctuation theorems \cite{PhysRevE.76.031132,PhysRevLett.104.090601}.
According to \cite{PhysRevLett.104.090601}, the total entropy production can be decomposed as follows
\begin{equation}
\Delta S_{tot} [i;\Lambda] = \Delta S_{na} [i;\Lambda] + \Delta S_{a} [i;\Lambda],
\end{equation}
\begin{align}
\label{na_path_entropy}
\textrm{where}\quad\Delta S_{na} [i;\Lambda] \equiv \ln \frac{\mathcal{P}[i;\Lambda]}{\bar{\mathcal{P}}^{+}[i;\Lambda]} \\
\label{a_path_entropy}
\Delta S_{a} [i;\Lambda] \equiv \ln \frac{\mathcal{P}[i;\Lambda]}{{\mathcal{P}}^{+}[i;\Lambda]}
\end{align}
Here, the symbol $+$ means dual (or time reversible) transition,
\begin{equation}
w^{+}_{ij}(\Lambda) \equiv \frac{p^{st}_{i}(\Lambda)}{p^{st}_{j}(\Lambda)} w_{ji}(\Lambda),
\end{equation}
and the forward and reversed path probabilities are given by
\begin{align*}
&{\mathcal{P}}^{+}[i;\Lambda] = \prod_{t=1}^{T}w^{+}_{i_{t}i_{t-1}}(\Lambda_{t}) p_{i_0}(0)\\
&\bar{\mathcal{P}}^{+}[i;\Lambda] = \prod_{t=1}^{T}w^{+}_{i_{t-1}i_{t}}(\Lambda_{t}) p_{i_T}(T),
\end{align*}
respectively.
Hence, (\ref{na_path_entropy}) and (\ref{a_path_entropy}) can be rewritten as
\begin{align*}
&\Delta S_{na} [i;\Lambda] = \ln \frac{p_{i_0}(0)}{p_{i_T}(T)} + \ln \prod_{t=1}^{T} \frac{p^{st}_{i_{t}}(\Lambda_{t}) }{p^{st}_{i_{t-1}}(\Lambda_{t}) } \\
&\Delta S_{a} [i;\Lambda] = \ln \prod_{t=1}^{T} \frac{w_{i_{t}i_{t-1}}(\Lambda_{t}) p^{st}_{i_{t-1}}(\Lambda_{t})}{w_{i_{t-1}i_{t}}(\Lambda_{t}) p^{st}_{i_{t}}(\Lambda_{t})}.
\end{align*}
Further, they are rewritten as
\begin{align}
\label{na_entropy_1}
&\Delta S_{na} [i;\Lambda] = \Delta S_{sys}[i;\Lambda] + \Delta S_{ex}[i;\Lambda]\\
\label{a_entropy_1}
&\Delta S_{a} [i;\Lambda] = \Delta S_{res}[i;\Lambda] - \Delta S_{ex}[i;\Lambda],
\end{align}
where
\begin{equation}
\label{ex_entropy}
\Delta S_{ex}[i;\Lambda] \equiv \ln \prod_{t=1}^{T} \frac{p^{st}_{i_{t}}(\Lambda_{t}) }{p^{st}_{i_{t-1}}(\Lambda_{t}) }
\end{equation}
\cite{PhysRevLett.104.090601}.
In the thermodynamic interpretations, $\Delta S_{ex} = - \beta Q_{ex}$ is the excess entropy production generated due to the excess heat,
and $\Delta S_{a} = -\beta Q_{hk}$ is the house-keeping entropy production due to the house-keeping heat \cite{oono1998steady,PhysRevLett.86.3463,PhysRevLett.104.090601,seifert2012stochastic,PhysRevE.60.R5017}.
If $\rho = \rho^{eq}$, then
\begin{equation}
\Delta S_{ex}[i;\lambda,\rho^{eq}] = \Delta S_{h}[i;\lambda,\rho^{eq}],
\end{equation}
since (\ref{same_with_equilibrium}), (\ref{Delta_S_epsilon-ex}) and (\ref{ex_entropy}).

By the path integrations of (\ref{total_path_entropy}), (\ref{na_path_entropy}) or (\ref{a_path_entropy}),
\begin{equation}
\sum_{\textrm{all } [i]}\mathcal{P}[i;\Lambda] e^{\Delta S_{tot/na/a} [i;\Lambda]} = \sum_{\textrm{all } [i]} \bar{\mathcal{P}}[i;\Lambda],
\end{equation}
we can obtain the three integral fluctuation theorems \cite{PhysRevE.76.031132},
\begin{equation}
\label{integral_FT}
\langle e^{-\Delta S_{tot/na/a} [\Lambda] }\rangle_{[i]} = 1
\end{equation}
where $\mean{\cdots}_{[i]}$ refers to average over all paths $[i]$.
According to \cite{PhysRevLett.104.090601}, they are essentially same with the three detailed fluctuation theorems,
\begin{equation}
\frac{P(\Delta S_{tot/na/a}[\Lambda])}{P(-\Delta S_{tot/na/a}[\Lambda])} = e^{\Delta S_{tot/na/a}[\Lambda]}
\end{equation}
where 
$\Delta S [\Lambda] \equiv \mean{ \Delta S [i;\Lambda]}_{[i]}$
is the average entropy change over all paths.
Applying the Jensen's inequality ($f(\mean{X}) \le \mean{f(X)}$ for a convex function $f(X)$ of a random variable $X$ \cite{ross1996stochastic}), the second law of the thermodynamics is derived,
\begin{equation}
\Delta S_{tot/na/a}[\Lambda] \ge 0,
\end{equation}
\cite{PhysRevE.76.031132,PhysRevLett.104.090601}.

According to \cite{PhysRevLett.104.090601}, the second law can be directly derived from (\ref{total_path_entropy}) by applying the Kullback-Leibler divergence,
\begin{equation}
\label{total_entropy_change}
\begin{split}
{\Delta S_{tot}[\Lambda]} 
= D\Big(\mathcal{P}[i;\Lambda] \Big| \bar{\mathcal{P}}[i;\Lambda] \Big) \ge 0.
\end{split}
\end{equation}
Here, $D(A_{i}|B_{i}) \equiv \sum_{i} A_{i}\ln\frac{A_{i}}{B_{i}} \ge 0$ is the Kullback-Leibler divergence, which is zero when $A_{i} = B_{i}$ for all states (or paths) $i$.
In the same manner, (\ref{na_path_entropy}) and (\ref{a_path_entropy}) yield
\begin{eqnarray}
\label{na_entropy_change}
{\Delta S_{na}[\Lambda]}
= D\Big(\mathcal{P}[i;\Lambda] \Big| \bar{\mathcal{P}}^{+}[i;\Lambda] \Big) \ge 0 \\
\label{a_entropy_change}
{\Delta S_{a}[\Lambda]}
= D\Big(\mathcal{P}[i;\Lambda] \Big| {\mathcal{P}}^{+}[i;\Lambda] \Big) \ge 0.
\end{eqnarray}
By the relations, the second law has been extended to $\Delta S_{na}[\Lambda]$ or $\Delta S_{a}[\Lambda]$ \cite{PhysRevE.76.031132,PhysRevLett.104.090601}.

\section{. Entropy Productions}
\label{sec:derivative}

\subsection{$\frac{d S_{sys}}{dt}$: System Entropy Production} 
The average system entropy change is given by
\begin{align}
\label{system_nonequilibrium_1}
& \Delta S_{sys} [\Lambda]  = \sum_{\textrm{all }[i]} \mathcal{P}[i;\Lambda]  \Delta S_{sys}[i;\Lambda] \\
&= \sum_{i_{T},\cdots,i_{0}} w_{i_{T}i_{T-1}}(\Lambda_T) \cdots w_{i_{1}i_{0}}(\Lambda_1) p_{i_{0}}(0)
\ln \frac{p_{i_0}(0)}{p_{i_{T}}(T)} \nonumber
\end{align}
In the limit $t_0 \rightarrow t-\Delta t$ and $t_{T} \rightarrow t$,
equation (\ref{system_nonequilibrium_1}) becomes
\begin{align}
\label{system_nonequilibrium_2}
& \frac{d S_{sys}}{dt} = \lim_{\Delta t \rightarrow 0 }
\frac{\sum_{ij} w_{ij}(\Lambda_t) p_{j}(t-\Delta t)
\ln \frac{p_{j}(t-\Delta t)}{p_{i}(t)} }{\Delta t} \nonumber \\
&= \lim_{\Delta t \rightarrow 0 } \frac{- \sum_{i} p_{i}(t) \ln {p_{i}(t)}
+ \sum_{j} p_{j}(t-\Delta t) \ln {p_{j}(t-\Delta t)} }{\Delta t} \nonumber \\
&= \lim_{\Delta t \rightarrow 0 } \frac{- \sum_{i} \Delta p_{i}(t) \ln {p_{i}(t - \Delta t)}}{\Delta t} \nonumber \\
&= - \sum_{i} \dot p_{i}(t) \ln {p_{i}(t)} 
\end{align}
where $\Delta p_i(t) = P_i(t) - p_i(t-\Delta t)$.
Since we have assumed that $p_{i}(t) = p^{eq}_{i}(\lambda)e^{-\theta_i -\phi_i}$ and $p^{eq}_{i}(\lambda)=e^{\beta(F-E_i)}$,
(\ref{system_nonequilibrium_2}) can be rewritten as 
\begin{align}
\label{derivative:S_eq}
\frac{d {S_{sys}} }{dt}
= \frac{d {S}}{dt}  + \frac{d  \theta}{dt} + \frac{d  \phi }{dt}
\end{align}
where ${S} \equiv -\beta F + \beta {E} $, $E \equiv \sum_{i}p_{i}(t){E_{i}}$ is the internal energy, and $F \equiv - \beta^{-1} \ln \sum_{i} e^{-\beta E_{i}}$ is the Helmholtz free energy.

\subsection{$\frac{d S_{h}}{dt}$, $\frac{d S_{\epsilon}}{dt}$ and $\frac{dS_{\nu}}{dt}$}
\label{Equilibrium_Excess_Entropy_Production}
In the same manner, from (\ref{Boltzmann_distribution}) and (\ref{Delta_S_epsilon-ex}), $\frac{d S_{h}}{dt}$ is calculated as follows
\begin{equation}
\label{my_excess_2}
  \begin{split}
    & \Delta {S_{h}} \equiv \sum_{ij} w_{ij}(\Lambda_{t})p_{j}(t-\Delta t) \ln \frac{p^{eq}_{i}(\lambda_{t})}{p^{eq}_{j}(\lambda_{t})} \\
    & = \sum_{i} p_{i}(t) \ln p^{eq}_{i}(\lambda_{t}) - \sum_{j} p_{j}(t-\Delta t) \ln p^{eq}_{j}(\lambda_{t}) \\
    & = \sum_{i} p_{i}(t) \left[ F - E_{i}\right] - \sum_{j} p_{j}(t-\Delta t) \left[ F - E_{i}\right] \\
    & = - \beta E ({t}) + \beta { E }({t-\Delta t}) \quad (\lambda \textrm{ is fixed.})
  \end{split}
\end{equation}
which is independent of $\rho$, and $F$ is canceled out in the calculations.
Therefore,
\begin{equation}
  \label{my_excess_1}
 \frac{d{S_{h}}}{dt} = -\beta \frac{\partial  E }{\partial t}.
\end{equation}
Here, 
\begin{equation}
\frac{ \partial  E }{\partial t}= \frac{d  E }{ d t} - \frac{\partial  E }{\partial \lambda}\dot \lambda = \frac{d  E }{ d t} -\frac{d  W }{ d t} = \frac{d Q }{dt}
\end{equation}
since the time derivative of the work done on the system is normally given by $\frac{d W }{ d t} = \frac{\partial   E }{\partial \lambda} \dot \lambda$ \cite{cohen2004note,jarzynski2004nonequilibrium}, and the first law of the thermodynamics, $\Delta {E} = W + Q$.
Therefore, (\ref{my_excess_1}) can be given by
\begin{equation}
\label{my_excess_entropy}
\frac{d { S_{h}}}{dt}  = -\beta \frac{d Q }{d t} 
\end{equation}

Hence, $\frac{d {S_{\epsilon}}}{dt}$ and $\frac{d {S_{\nu}}}{dt}$ are given by
\begin{equation}
\label{epsilon_derivative}
\begin{split}
\frac{d {S_{\epsilon}}}{dt} &= \frac{d {S_{sys}}}{dt} + \frac{d {S_{h}}}{dt} \\
&=\frac{d {S} }{ d t } + \frac{d {\theta}}{dt} + \frac{d {\phi}}{dt} - \beta \frac{d Q }{d t}
\end{split}
\end{equation}
and
\begin{equation}
\label{nu_derivative}
\begin{split}
\frac{d {S_{\nu}}}{dt} &= \frac{d {S_{res}}}{dt} - \frac{d {S_{h}}}{dt} = \frac{d {S_{res}}}{dt} + \beta \frac{d Q }{d t},
\end{split}
\end{equation}
respectively.

\subsection{$\frac{d S_{ex}}{dt}$, $\frac{d S_{na}}{dt}$ and $\frac{d S_{a}}{dt}$}
In the same manner, replacing $t_0 \rightarrow t-\Delta t$ and $t_T \rightarrow t$, (\ref{ex_entropy}) yields that
\begin{equation}
  \label{excess_derivative}
  \begin{split}
    & \Delta{{S_{ex} (\Lambda)}} \equiv \sum_{ij} w_{ij}(\Lambda)p_{j}(t-\Delta t) \ln \frac{p^{st}_{i}(\Lambda)}{p^{st}_{j}(\Lambda)} \\
    &=\sum_{ij} w_{ij}(\Lambda)p_{j}(t-\Delta t) \left[ \ln \frac{p^{eq}_{i}(\lambda)}{p^{eq}_{j}(\lambda)} - \theta_{i} + \theta_{j}\right]\\
    & \rightarrow \frac{d{S_{h} (\Lambda)}}{dt} - \frac{\partial {\theta}}{\partial t}
    \quad \left(\textrm{for fixed }\Lambda \right)
  \end{split}
\end{equation}
since (\ref{Boltzmann_distribution}), (\ref{steady_distribution}), (\ref{my_excess_2}) and (\ref{my_excess_1}).
Therefore,
\begin{align}
\label{na_derivative}
&\frac{d {S_{na}}}{dt} = \frac{d {S_{sys}}}{dt} + \frac{d {S_{ex}}}{dt}\\
&\quad= \frac{d {S_{sys}}}{dt} + \frac{d{S_{h} (\Lambda)}}{dt} - \frac{\partial {\theta}}{\partial t}
= \frac{d {S_{\epsilon}}}{dt} - \frac{\partial {\theta}}{\partial t} \ge 0 \nonumber
\end{align}
and
\begin{align}
\label{a_derivative}
&\frac{d {S_{a}}}{dt} = \frac{d {S_{res}}}{dt} - \frac{d {S_{ex}}}{dt}\\
&= \frac{d {S_{res}}}{dt} - \frac{d{S_{h} (\Lambda)}}{dt} + \frac{\partial {\theta}}{\partial t} 
= \frac{d {S_{\nu}}}{dt} + \frac{\partial {\theta}}{\partial t} \ge 0.\nonumber
\end{align}

\subsection{$\frac{\partial F}{\partial t} =  \frac{\partial E}{\partial t}|_{eq}$}
\label{F_E_eq}

\begin{align}
&\frac{\partial F}{\partial t} 
= - \beta^{-1} \frac{\partial \ln \sum_{i}e^{-\beta E_{i}} }{\partial t}
= -\beta^{-1}\sum_{i} \frac{-\beta \frac{\partial E_{i}}{\partial t} e^{-\beta E_{i}}}{\sum_{i'}e^{-\beta E_{i'}}} \nonumber \\
&= \sum_{i} \frac{\partial E_{i}}{\partial t} e^{\beta(F-E_{i})}
= \sum_{i} \frac{\partial E_{i}}{\partial t} p^{eq}_{i}(\lambda) 
= \frac{\partial E}{\partial t}\bigg|_{eq}
\end{align}

\section{. Negative Entropy Production}
\label{derive:negative_entropy}
In the out-of-equilibrium processes, the average of (\ref{symmetric_entropy}) is calculated as follows
\begin{align}
  \label{der:negative_epsilon_1}
  &\Delta S_{\epsilon} [\Lambda_{out}] = \sum_{\textrm{all }[i]} \mathcal{P}[\Lambda] \Delta S_{\epsilon}[i;\Lambda]
  \nonumber \\
  &= \sum_{i_T,\cdots,i_0} w_{i_Ti_{T-1}}(\Lambda_T) \cdots w_{i_1i_0}(\Lambda_0) p_{i_0}(0) \nonumber \\
  &\quad\times\ln \prod_{t=1}^{T}
  \frac{\epsilon_{i_ti_{t-1}}(\lambda)p_{i_0}(0)}{\epsilon_{i_{t-1}i_{t}}(\lambda)p_{i_{T}}(T)} \nonumber \\
  &= \sum_{i_T,\cdots,i_0} w_{i_Ti_{T-1}}(\Lambda_T) \cdots w_{i_1i_0}(\Lambda_0) p_{i_0}(0) \nonumber \\
  &\quad\times\ln \prod_{t=1}^{T}
  \frac{p^{eq}_{i_t}(\lambda)p_{i_0}(0)}{p^{eq}_{i_{t-1}}(\lambda)p_{i_{T}}(T)}
\end{align}
since $\epsilon_{ij}(\lambda) /  \epsilon_{ji}(\lambda) = p^{eq}_{i}(\lambda) / p^{eq}_{j}(\lambda)$.
Therefore,
\begin{align}
  &(\ref{der:negative_epsilon_1})= \sum_{i_T,\cdots,i_0} w_{i_Ti_{T-1}}(\Lambda_T) \cdots w_{i_1i_0}(\Lambda_0) p_{i_0}(0) \nonumber \\
  &\quad\times \left[\ln \frac{p^{eq}_{i_T}(\lambda)}{p_{i_{T}}(T)} + \ln \prod_{t=1}^{T}\frac{p^{eq}_{i_{t-1}}(\lambda)}{p^{eq}_{i_{t-1}}(\lambda)} \right] \nonumber \\
  &= \sum_{i_T,\cdots,i_0} w_{i_Ti_{T-1}}(\Lambda_T) \cdots w_{i_1i_0}(\Lambda_0) p_{i_0}(0) \ln \frac{p^{eq}_{i_T}(\lambda_T)}{p_{i_{T}}(T)} \nonumber \\
  &= \sum_{i_T} p_{i_T}(T) \ln \frac{p^{eq}_{i_T}(\lambda)}{p_{i_{T}}(T)} \nonumber \\
  &= - D \Big( p^{st}_{i}(\Lambda_T) \Big| p^{eq}_{i}(\lambda) \Big) \le 0 
\end{align}
since (\ref{assume_out}).



\end{appendix}

\bibliographystyle{apsrev4-1}

\bibliography{log_decomposition}

\end{document}